\documentclass[prl,twocolumn,floatfix,superscriptaddress,aps]{revtex4-2}

\usepackage[utf8]{inputenc}
\usepackage[T1]{fontenc}

\usepackage{graphicx}
\usepackage{bm}

\usepackage{amsmath}
\usepackage{units}

\usepackage{hyperref}
\hypersetup{colorlinks=true, citecolor=blue, urlcolor=blue, linkcolor=blue}

\begin{document}
\title{Time-resolved electron dynamics in antiferromagnetic CoO(001) thin films}

\author{Mathias Augstein}
\author{Friederike Wührl}
\author{Konrad Gillmeister}
\affiliation{Institute of Physics, Martin Luther University Halle-Wittenberg, D-06099 Halle (Saale), Germany}

\author{Cheng-Tien Chiang}
\affiliation{Institute of Atomic and Molecular Sciences, Academia Sinica, Taipei, Taiwan}

\author{Wolf Widdra}
\affiliation{Institute of Physics, Martin Luther University Halle-Wittenberg, D-06099 Halle (Saale), Germany}

\email{wolf.widdra@physik.uni-halle.de}

\date{\today}

\begin{abstract}
Ultrathin antiferromagnetic CoO(001)-(1x1) films of 2 and 4 monolayers (ML) on Ag(001) are investigated by time- and angle-resolved two-photon photoelectron (2PPE) spectroscopy. Pump-probe spectra show unoccupied states between 3.6 and 4.0\,eV above the Fermi level ($E_{F}$), which are identified as image potential states with momentum-dependent lifetimes between 12 to 20\,fs. Electrons photoexcited across the band gap show lifetimes of 27\,fs at the bottom of the Co $3d$ $t_{2g}$-derived conduction band minimum. This lifetime is much shorter than in conventional semiconductors. Our observations point either to strong electron correlation effects as has been demonstrated for NiO or to an ultrafast relaxation pathway via metallic substrate states.  
\end{abstract}


\maketitle

\paragraph{Introduction} 

Among the transition metal oxides, the antiferromagnetic (AFM) oxides form a family of strongly correlated electron systems, where the local electron correlations lead to insulating behavior and simultaneously stabilize the long-range magnetic ordering via Anderson’s superexchange mechanism \cite{Anderson50}. Paradigmatic examples are Mott and charge-transfer (CT) insulators, whose optical properties are determined by the charge transfer between the oxygen ligand and the correlated orbital (typically  $d$) states \cite{Sawatzky84}. CoO and NiO belong to the charge-transfer insulators. Whereas NiO has fully occupied  $3d$ $t_{2g}$ states and an $e_g$-derived conduction band (upper Hubbard band), the lowest unoccupied state has $t_{2g}$ character in CoO \cite{Sawatzky84,Schroen15}. Calculations yield charge-transfer band gaps between 2.3 and 3.6\,eV \cite{Das15,Roedel09,Tran06,Jiang10,Dalverny10}, whereas experimentally values have been observed between 2.5 and 2.7\,eV \cite{Otto16,Pratt59,Powell70}. The occupied electronic structure of both these oxides have been intensively studied over the past decades and provided insights into the importance of electron-electron interaction beyond one-electron approximation \cite{Shen_PRB90,Huefner_AdvPhys94}.

Conventionally a long lifetime beyond 100\,fs for photoexcited electrons at the bottom of the conduction band in semiconductors with a few electron volts band gaps has been observed \cite{Haight_SSR95}. In contrast, ultrashort lifetimes in the range of 10\,fs have been observed in the strongly correlated charge-transfer insulator NiO \cite{Gillmeister20}. Here the photoexcitation across the charge-transfer band gap leads to an ultrafast coupled charge and spin dynamics into a many-body excitation \cite{Gillmeister20}. The latter has been experimentally identified as a many-body in-gap state with long-lived oscillations due to the superexchange interaction between antiferromagnetically aligned next-nearest neighboring Ni ions \cite{Fischer09}. The frequency of these oscillations corresponds to the strength of the magnetic superexchange in NiO, whereas the oscillations vanish slightly above the Néel temperature \cite{Gillmeister20}. However, time-resolved photoemission studies of other transition metal oxides are rather scarce \cite{Aeschlimann_SS25}.

In this work, we investigate photoexcited transient states in CoO(001) thin films using time-resolved two-photon photoemission (2PPE) spectroscopy with two independently tunable ultraviolet (UV) light sources. Our focus is on tracking the dynamics of excited electrons in the Co $3d$ conduction band and in the image potential states of the transition-metal oxide CoO. Thin films of CoO are grown on a Ag(001) substrate to avoid charging in electron spectroscopy methods. These thin films have been characterized previously by scanning tunneling microscopy (STM), diffraction methods, as well as angle-resolved photoelectron spectroscopy \cite{Hagendorf03,Sindhu04,Cheng08,Barman18}. Although pioneering 2PPE studies of the unoccupied states in CoO films on Ir(100) have been performed previously and the conduction band minimums at 2\,eV above the Fermi level ($E_{F}$) for thicker layers and at 2.13\,eV for 4 monolayers (ML) were identified \cite{Otto16}, no time-resolved 2PPE studies have yet been reported for CoO thin films so far.

\paragraph{Growth and antiferromagnetic order}

\begin{figure}
  \centering
  \includegraphics[width = 0.8\columnwidth]{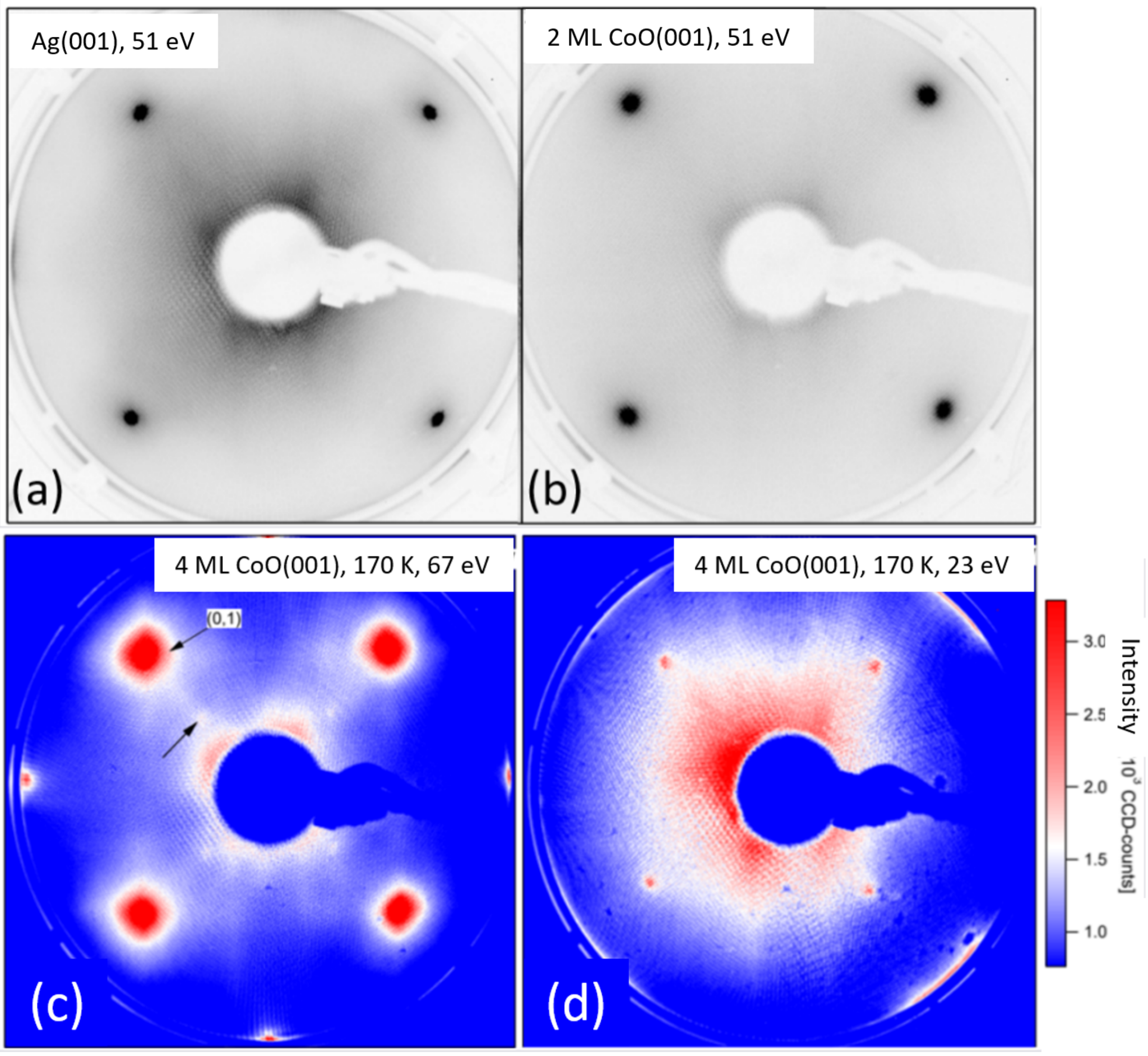}
  \caption{LEED structure of CoO(001) thin films on Ag(001). Prior to growth (a) and upon deposition of 2 ML CoO (b) at a kinetic energy of \unit[51]{eV}. The four first-order diffraction spots are slightly broadened upon CoO deposition. (c,d) Diffraction at \unit[170]{K} for a film thickness of 4 ML. First and half-order spot are marked by arrows. The latter, which are due to the antiferromagnetic (2x1) superstructure, are better visible in (d) for a lowered kinetic energy of \unit[23]{eV}. }
  \label{fig:LEED}
\end{figure}

\begin{figure}
  \centering
  \includegraphics[width = 0.8\columnwidth]{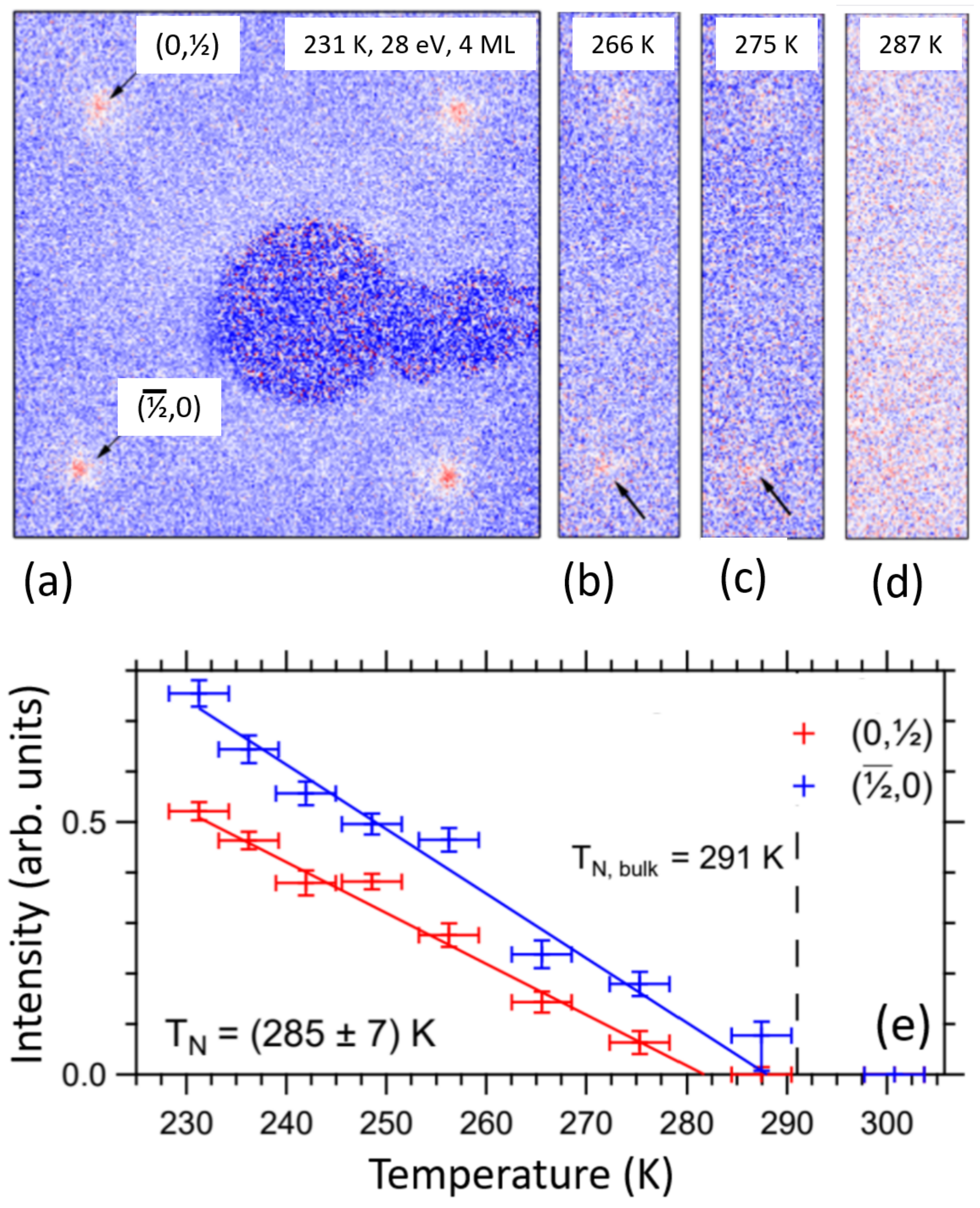}
  \caption{Antiferromagnetic (2x1) diffraction peaks of 4 ML CoO(001)/Ag(001) for various surface temperatures of 
  231, 266, 275, and 287 K in (a), (b), (c), and (d), respectively. (e) Half-order diffraction intensities as function of substrate temperature. The antiferromagnetic order vanishes at about \unit[285$\pm$5]{K}.}
  \label{fig:LEED_temp}
\end{figure}

The CoO thin films in this study have been grown on Ag(001) by thermal evaporation of Co from a crucible in an oxygen atmosphere of 1$\times$\,10$^{-6}$\,mbar. The growth has been controlled by X-ray photoelectron spectroscopy (XPS) and low-energy electron diffraction (LEED). The oxide thicknesses have been determined by analysis of Co and O intensities in comparison to the attenuation of the Ag substrate signal in XPS (not shown here). Accordingly, CoO film thicknesses of 1.9 and 3.8\,ML were finally assigned, which we will refer to here as 2 and 4\,ML for simplicity. Due to the lattice mismatch of CoO and Ag of about \unit[4]{\%} \cite{Schindler_SS09}, the epitaxial CoO(001) thin films grow slightly strained with the (1$\times$1) substrate superstructure as depicted in Fig.\,\ref{fig:LEED}. In comparison to the (1x1) spots of the bare Ag(001) substrate (shown in Fig. \ref{fig:LEED}(a)), the spots for \unit[2]{ML} CoO are slightly widened as shown in Fig.\,\ref{fig:LEED} (b). At \unit[4]{ML} CoO thickness and T\,=\,\unit[170]{K}, weak half order spots are discernible in (c) due to antiferromagnetic (2x1) and (1x2) domains (marked as arrow). At lower kinetic electron energies these antiferromagnetic diffraction peaks gain intensity as shown in Fig.\,\ref{fig:LEED}(d) for \unit[23]{eV}. To improve the contrast, the diffraction patterns in Fig.\,\ref{fig:LEED_temp} are normalized to the pattern at \unit[300]{K} in the absence of the half-order spots. Figs.\,\ref{fig:LEED_temp}(a)-(d) show the LEED pattern for 287, 275, 266, and 231\,K. The clear half-order spots, which are visible at 231\,K, vanish with increasing temperature and are no longer detectable at 287\,K. These temperature-dependent LEED experiments are summarized in Fig.\,\ref{fig:LEED_temp}(e) for the two half-order spots (-\textonehalf,0) and (0, \textonehalf). The intensity of the additional antiferromagnetic half-order diffraction vanishes at about \unit[285$\pm$5]{K}, which is slightly lower as compared to the bulk CoO Néel temperature ($T_{N}$) of \unit[291]{K}.These data exhibit clearly the antiferromagnetic ordering of the 4\,ML CoO(001) thin film, even in the presence of a strained growth on Ag(001). Our observation of antiferromagnetic 4\,ML CoO is consistent with its strong thickness dependent $T_{N}$ \cite{Xu_NJP20,Barman_JMMM20,Ambrose_PRL96}, but in our case $T_{N}$ remains below the bulk value.

\paragraph{Image potential states}

\begin{figure}
  \centering
  \includegraphics[width = 0.6\columnwidth]{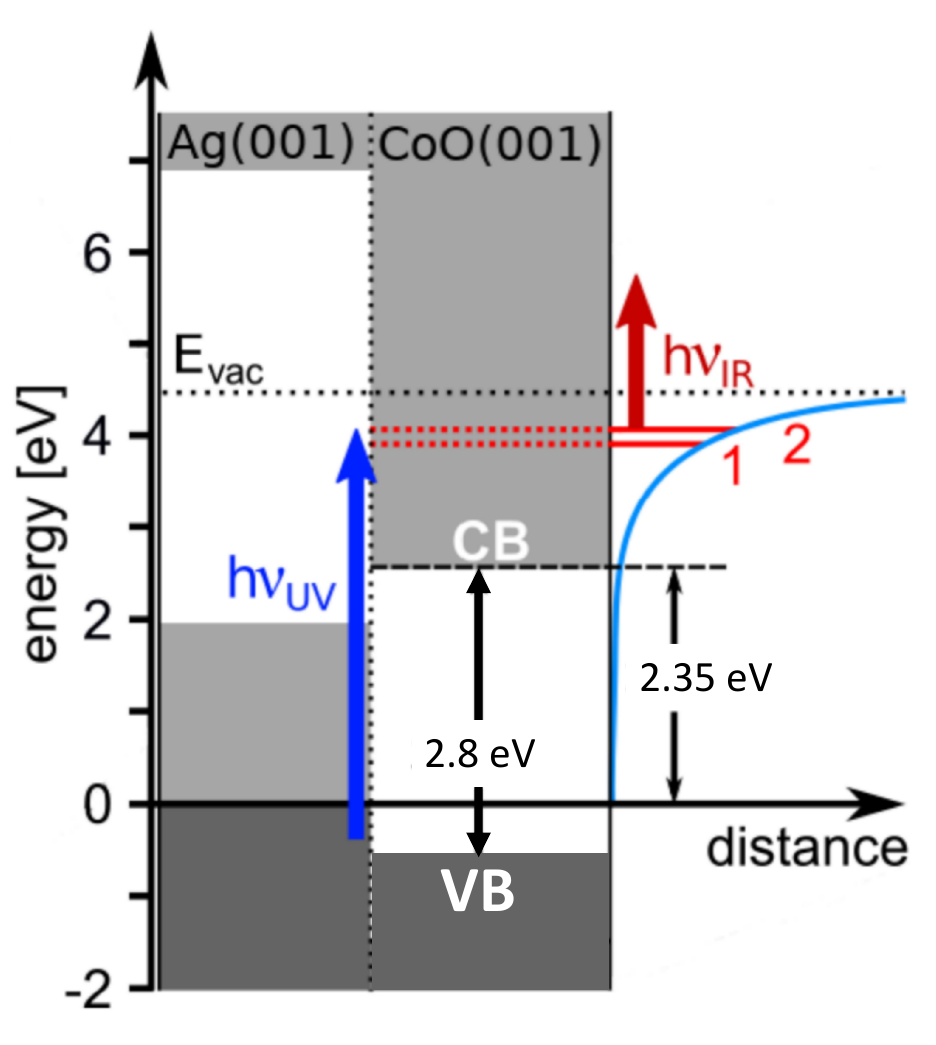}
  \caption{Schematic energy level diagram of the CoO(001)/Ag(001) interface with O2p-derived valence band (VB) and Co3d-derived conduction band (CB). In front of the CoO surface the image potential is indicated (blue line) with two image states (red) below the vacuum level $E_{vac}$.}
  \label{fig:BandStruc}
\end{figure}

In the following, we will address the image potential (IP) states on the CoO(001) surface by time-resolved 2PPE experiments. The relevant energy scheme is depicted in Fig.\,\ref{fig:BandStruc}. An ultrashort ultraviolet (UV) laser pulse excites electrons from close to $E_{F}$ in the Ag(001) substrate into unoccupied states of the CoO(001) thin film. A second infrared (IR) laser pulse photo-ionizes the transiently populated excited states, as shown for the second IP states in Fig.\,\ref{fig:BandStruc}. In this relevant energy-momentum region of unoccupied states around $k_{\parallel}=0$, the Ag(001) substrate has a directional band gap opening at \unit[2]{eV} above $E_{F}$ \cite{Gillmeister18}.

\begin{figure}
  \centering
  \includegraphics[width = 0.9\columnwidth]{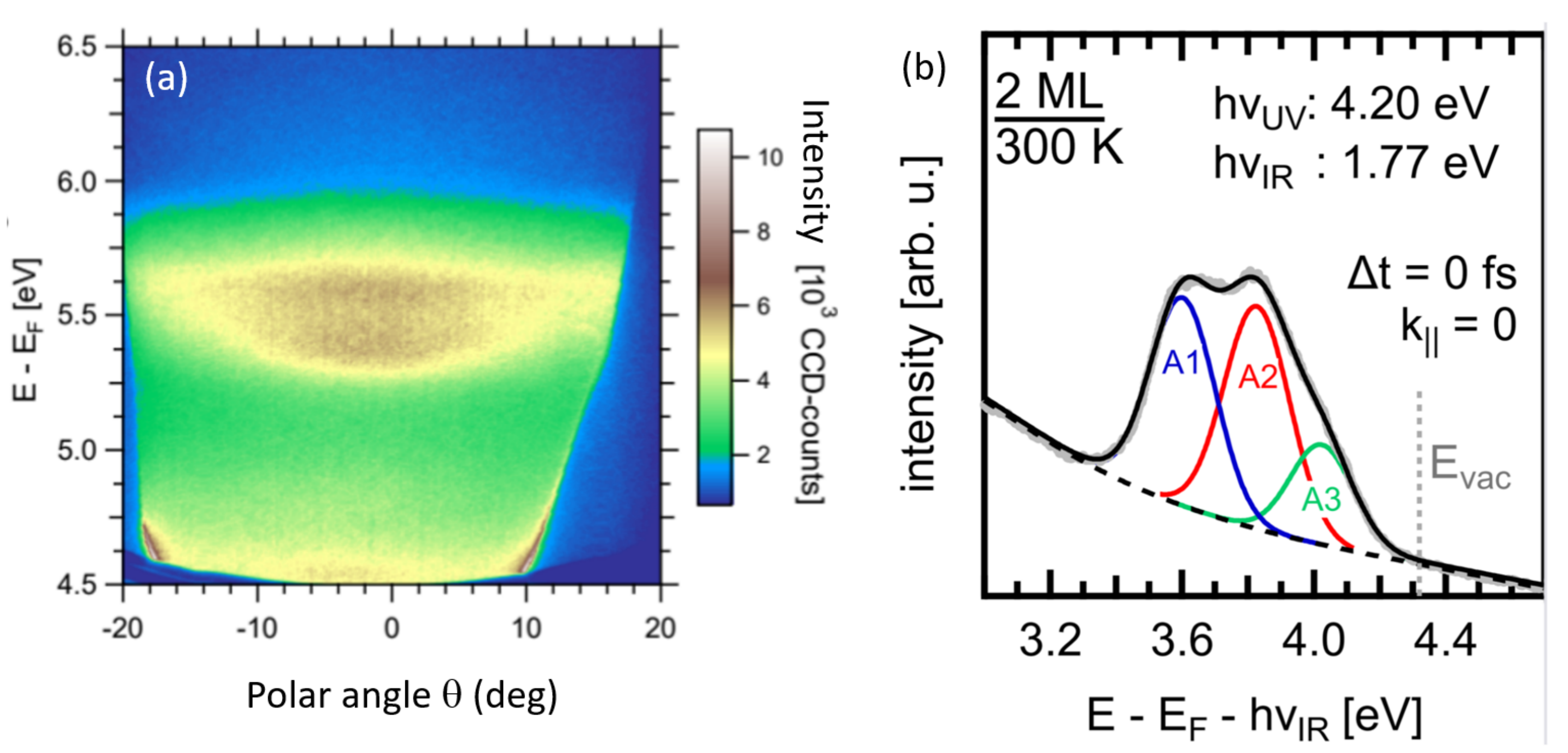}
  \caption{2PPE of the image potential states A1, A2, and A3 just below the vacuum energy for a 2ML thick CoO(001) thin film grown on Ag(001), populated with 4.20 eV and probed with 1.77 eV photons.The spectrum is plotted as function of intermediate-state energy relative to the Fermi level $E_F$ (a) }
  \label{fig:Imag1b}
\end{figure}

Figure\,\ref{fig:Imag1b} shows the 2PPE data for a \unit[2]{ML} thick CoO(001) thin film using pump-probe photon energies of 4.2 and 1.77\,eV, respectively, with zero time delay in between. The angle-dependent data in Fig.\,\ref{fig:Imag1b}(a) display two clear close-lying peaks with a band minimum at $k_\parallel=0$ and their parabolic dispersion towards higher emission angles. The effective mass of the lower energy peak A1 could be estimated as m$_{eff}$\,$\approx$\,0.8\,$\pm$\,0.4\,m$_{e}$, with m$_{e}$ being the free electron mass. The photoelectron spectrum in (b) emphasizes the two peaks A1 and A2 as well as an additional shoulder A3 at 3.6, 3.83, and 4.1\,eV, respectively, close to the vacuum level at 4.3\,eV. We assign these three states to the first IP states in front of the bilayer CoO(001) based on their energetic position just below the vacuum energy $E_{vac}$. 

\begin{figure}
  \centering
  \includegraphics[width = 0.99\columnwidth]{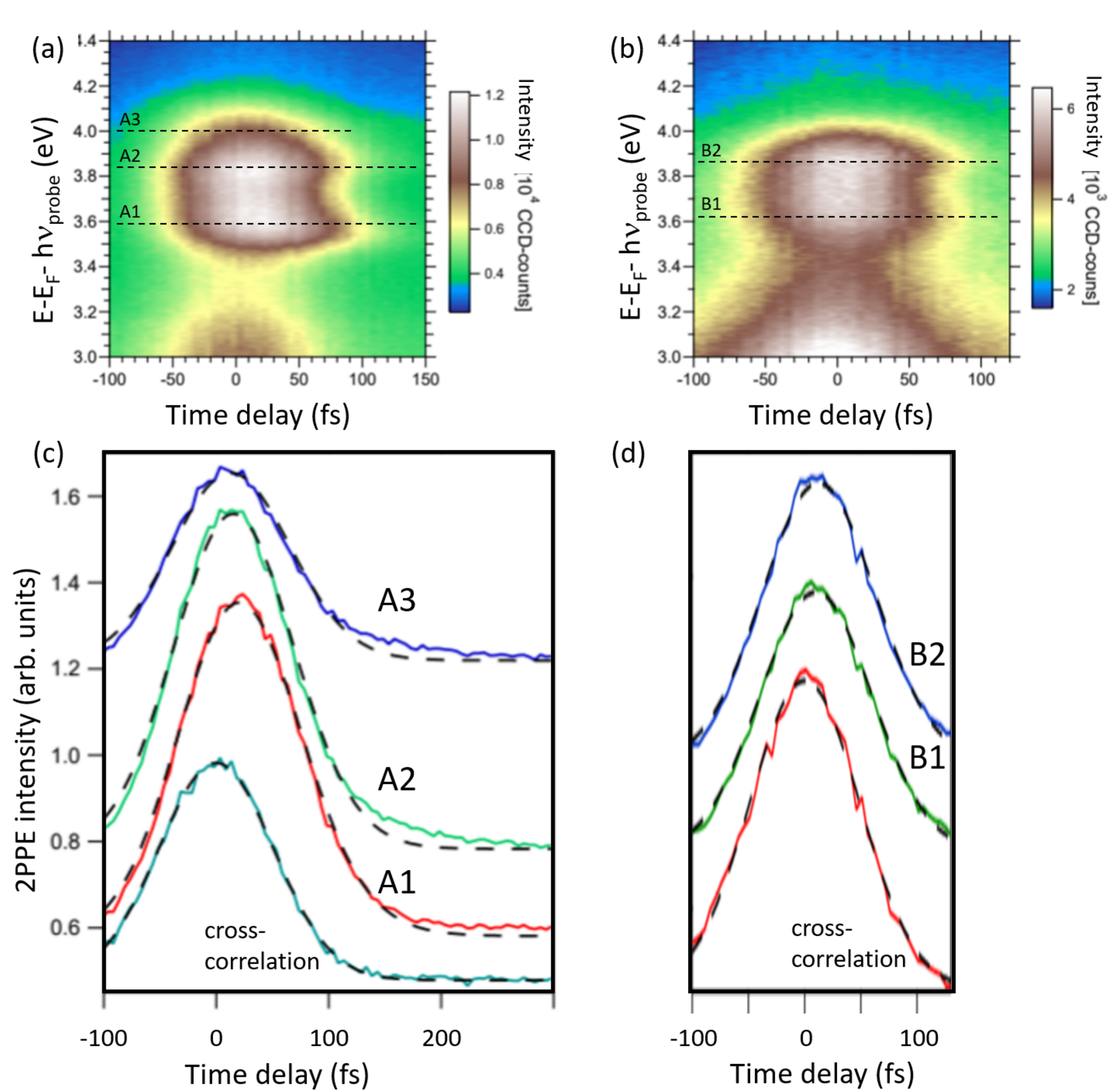}
  \caption{Time-resolved 2PPE spectra of the first image potential states for (a) 2\,ML thick and (b) 4\,ML thick CoO(001) thin films on Ag(001) at $k_\parallel=0$. (a) and (c): Photon energies of 1.77 and \unit[4.20]{eV}. (b) and (d): Photon energies of 1.70 and \unit[4.12]{eV}.}
  \label{fig:ImagPotLifetimes}
\end{figure}

The lifetime of the unoccupied electronic states A1, A2, and A3 can be directly quantified via time-resolved 2PPE at T\,=\,\unit[170]{K} as shown in Fig.\,\ref{fig:ImagPotLifetimes}(a) and (b), respectively, for photon energies of \unit[1.77]{eV} and \unit[4.20]{eV} with a cross correlation of $120\pm5$ and $115\pm5$\,fs between both. The data in (a) present the dynamics of the two IP states A1 and A2 at 3.6 and 3.83\,eV for 2\,ML CoO(001) as discussed in Fig.\,\ref{fig:Imag1b}. The slightly slower decay of their 2PPE intensities at positive pump-probe delays confirms their transient population triggered upon the UV excitation and subsequent photo-ionization by the IR pulses. Quantitative fits for the lifetimes (dashed black lines) are depicted in Fig.\,\ref{fig:ImagPotLifetimes}(c) for the three IP states. For comparison, the cross correlation traces (bottom curve in Fig.\,\ref{fig:ImagPotLifetimes}(c) and (d)) are also given as extracted from the photoemission data at an intermediate state energy of \unit[2.8]{eV}. The small shift of the maxima of the IP spectra to larger $\Delta t$ with respect to the cross correlation (bottom curve) as well as the slightly asymmetric line shape can be well fitted with lifetimes of the IP state of $23\pm5$, $16\pm5$, and $14\pm8$\,fs for the states A1, A2, and A3, respectively. Within the uncertainty, all three IP states have ultrashort lifetimes between 14 and 25\,fs. 

Figures\,\ref{fig:ImagPotLifetimes}(b) and (d) show the time-resolved 2PPE data for 4\,ML CoO(001) on Ag(001) at $k_\parallel =0$, where the data reveal the two image potential states, B1 at 3.62 and B2 at 3.87\,eV. The photoemission intensity for different pump-probe delays in \ref{fig:ImagPotLifetimes}(d) exhibits slightly asymmetric line shapes and a shift of the peak maximum to larger $\Delta t$ as compared to the cross correlation (bottom curve). Both observations indicate finite lifetimes of the states B1 and B2, which we estimate to $10\pm5$ and $13\pm5$\,fs, respectively. An finite coupling of the IP states to the unoccupied CoO states might be an explanation for the ultrashort lifetime seen for 2 and 4\,ML CoO(001) thin films. Such coupling is consistent with the strongly thickness-dependent field emission resonances in the range of 3.5 to 5\,eV in scanning tunneling spectroscopy of the very same system \cite{Grosser_Diss08,Hagendorf03}, where the first resonance with a full-width-at-half-maximum of about 0.3\,eV has been observed.

\paragraph{Lifetime at the bottom of the conduction band}

\begin{figure}
  \centering
  \includegraphics[width = 0.8\columnwidth]{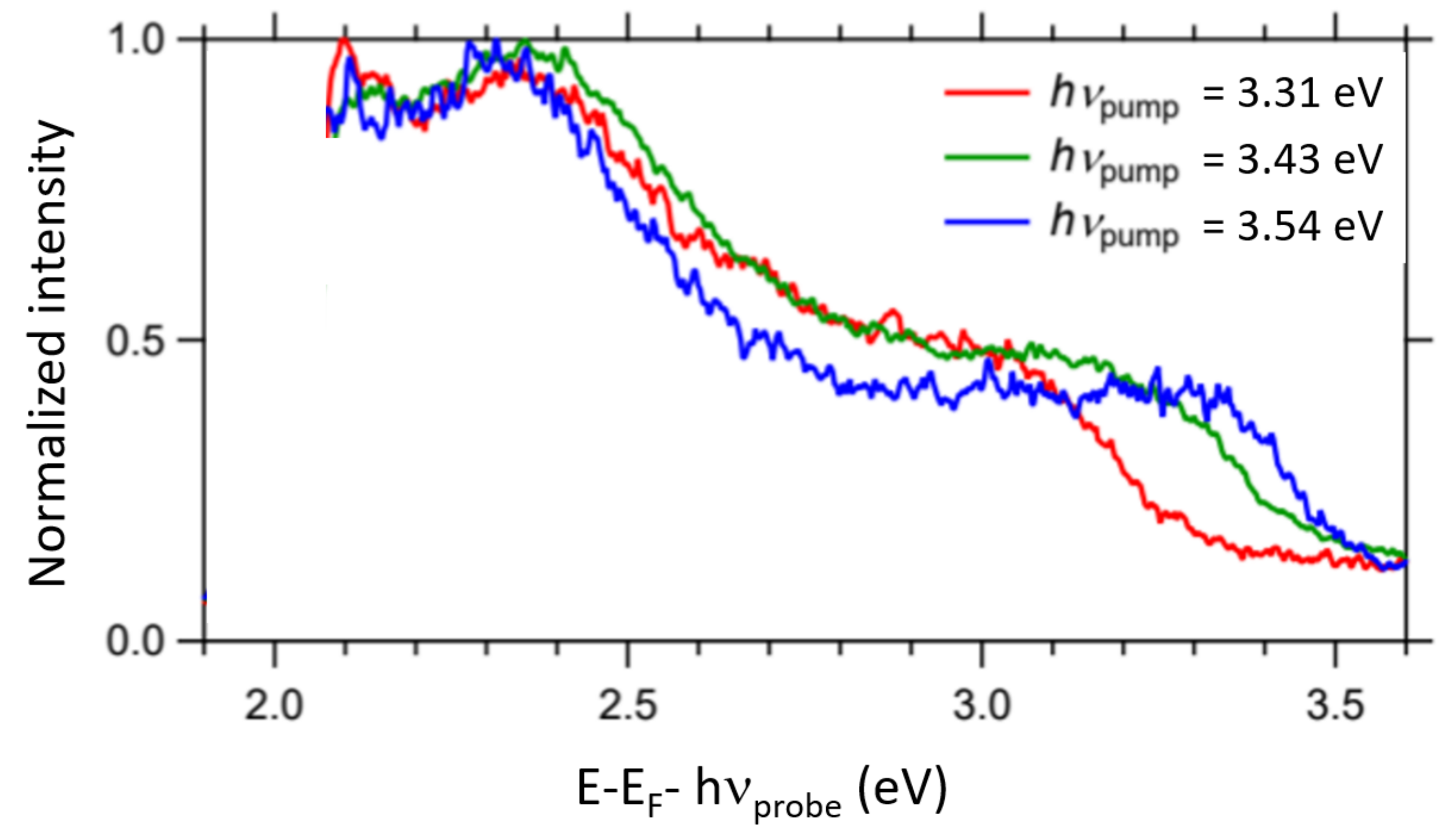}
  \caption{2PPE spectra for three different pump energies around the conduction band minimum for a 2\,ML thick CoO(001) thin film on Ag(001). probe energy 2.07\,eV and variable pump energies of 3.31, 3.43, and 3.54\,eV.}
  \label{fig:Spectrum_CBM_2ML}
\end{figure}

\begin{figure}
  \centering
  \includegraphics[width = 0.8\columnwidth]{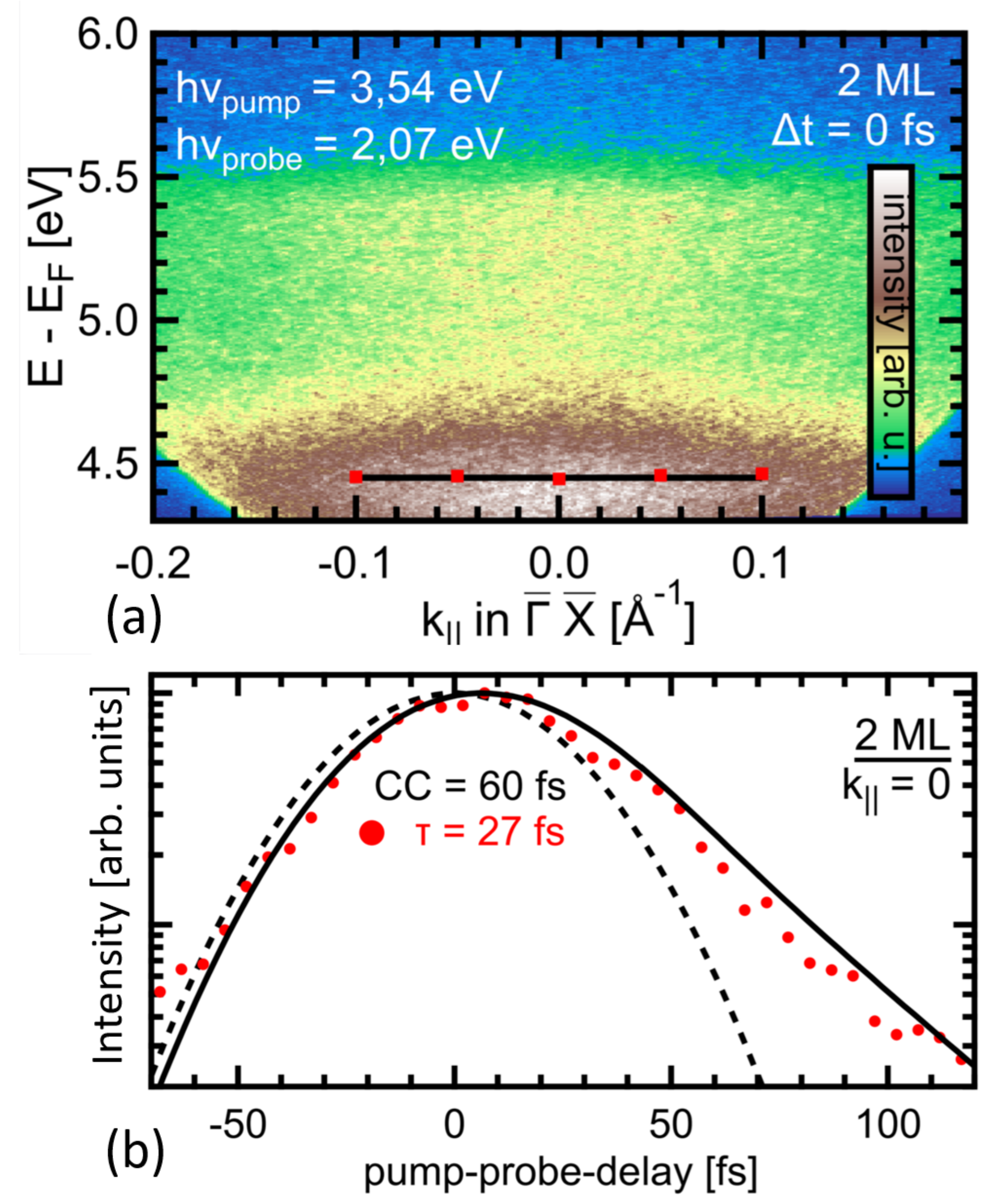}
  \caption{(a) Momentum-resolved 2PPE spectra around the conduction band minimum for a 2\,ML thick CoO(001) thin film on Ag(001). (b) Intensity at the conduction band minimum as function of pump-probe delay. Photon energies of 3.54\,and 2.07\,eV have been used to pump and probe the state.}
  \label{fig:CBM_2ML}
\end{figure}

Electrons photoexcited across the band gap of CoO can further reveal important information about its conduction band and provide a direct comparison to recent time-resolved data for excitation into the conduction band of NiO(001) as reported by Gillmeister \textit{et al.} \cite{Gillmeister20}. Figure\,\ref{fig:Spectrum_CBM_2ML} shows 2PPE data for a constant probe energy of 2.07\,eV and variable pump energies of 3.31, 3.43, and 3.54\,eV for 2\,ML CoO(001) on Ag(001). All of these three pump photon energies are in the order of the CoO band gap. On the intermediate state energy scale, the three spectra show a common peak around \unit[2.35]{eV} above $E_{F}$, which is assigned to the Co $3d$ $t_{2g}$-derived conduction band. This observation is in good agreement with the reported position of the conduction band at \unit[2.13]{eV} for 4\,ML CoO thin films on Ir(100) \cite{Otto16} and the value of 1.8\,$\pm0.3$\,eV above $E_{F}$ for CoO bulk \cite{vanElp91}. The high-energy cut-off in Fig.\,\ref{fig:Spectrum_CBM_2ML} is assigned to photoemission from $E_F$ of the underlying Ag(001), which shifts with the pump photon energy on this intermediate state energy scale. Momentum-resolved data in Fig.\,\ref{fig:CBM_2ML}(a) indicate a flat dispersion in the $\overline{\Gamma X}$ direction up to 0.13\,$\text{\AA}^{-1}$ as is expected for the Co $3d$ $t_{2g}$-derived conduction band bottom \cite{Schroen15}. Time-resolved data for this state at $k_\parallel=0$ are depicted in Fig.\,\ref{fig:CBM_2ML}(b) on a logarithmic intensity scale. There the 2PPE data (red circles) are compared to the pump-probe cross correlation (black dashed line). Whereas the latter is symmetric around $\Delta t=0$, the 2PPE data show a broader exponential decay for positive delays as well as a small shift of its maximum also to positive delays. Both can be well fitted by a lifetime of \unit[27]{fs} as indicated by the black solid line. In view of the band gap of around \unit[2.7]{eV}, this short lifetime is distinctly different from the picosecond lifetime at the conduction band minimum of conventional semiconductors \cite{Haight_SSR95}, and it is consistent with the observation on \textit{thicker} NiO films with their sub-\unit[10]{fs} lifetime of the Ni $3d$-derived conduction band \cite{Gillmeister20}.

\paragraph{Comparison with NiO and calculated density of states}

The observed lifetime above for IP states on CoO(001) are substantially shorter than the values for NiO(001) thin films with comparable thicknesses on the same substrate \cite{Gillmeister18}. This observation is consistent with the scanning tunneling spectroscopy (STS) experiments \cite{Grosser_Diss08}, showing the spectral width on CoO generally broader than that on the NiO by a factor of about 2. The full-width-at-half-maximum of the first field emission resonances in STS between 3 to 4\,eV, which can provide an orders-of-magnitude estimation for the lifetime of the first IP state, is 0.29 and 0.32\,eV on the 2 and 4\,ML CoO films, significantly broader than the 0.17 and 0.18\,eV width on the comparable NiO films. 

The conduction band feature in STS on CoO at around 2 to 3\,eV has a half width of about 0.28 and 0.31\,eV on the 2 and 4\,ML films \cite{Shantyr_ThinSolidFilm04,Grosser_Diss08}. For comparison, the half width amounts to 0.13 and 0.17\,eV on NiO of the same thicknesses. Since the exchange coupling is about 30\,\% weaker in CoO than NiO and the highest magnon frequency in CoO is about three times smaller than that in NiO \cite{Guntram_PRB09}, a magnetic origin or its coupling with phonons \cite{Kant_PRL12} would be tentatively excluded for the shorter electron lifetime at the conduction band bottom observed here for CoO. 

One possible explanation for the shorter lifetime in CoO than NiO would be due to the 4\,\% lattice mismatch in CoO on Ag(001) which is larger than the 2\,\% for NiO on Ag(001) \cite{Schindler_SS09,Giovanardi_ThinSolidFilm03}. As a consequence, the relaxation of CoO films starts already at around 3\,ML via the formation of dislocations \cite{Torelli_SS07}, in contrast to the onset of relaxation at 10\,ML for NiO \cite{Giovanardi_ThinSolidFilm03}. The earlier onset of strain relaxation in CoO than NiO, therefore, may lead to more structural imperfection when comparing both films of the same thickness below 10\,ML, explaining tentatively the shorter lifetime in CoO than NiO via electron-defect scattering \cite{Echenique_SSR04}.

Another possible explanation for the shorter electron lifetime in CoO would be its different electronic occupation compared to NiO, since their effective strength of electron-electron interaction are comparable \cite{OlaldeVelasco_PRB11}. Because of the $d^{7}$ configuration of the Co$^{2+}$ ion with one less $d$ electron than the $d^{8}$ of the Ni$^{2+}$ ion, there are more many-electron configurations of the Co$^{2+}$ which could potentially lead to more $d$-$d$ excitation channels within the optical band gap \cite{Abdallah_PRB24,Wray_PRB13,Cowley_PRB13,Chiuzbaian_PRB08}. These excitation channels may provide energy and momentum dissipation pathways for the optically excited electrons in our time-resolved 2PPE experiments \cite{Gillmeister18}. Despite that quantitative evaluation of these many-body dissipation pathways would still be challenging \cite{laTorre_RMP21,Kemper_PRX18}, qualitatively the more available empty $d$ states for the excited electrons in our 2PPE experiments could be inferred from the calculated density of states \cite{Das15}, where CoO possesses more states in the range of 3 to 5\,eV above $E_{F}$ as compared to NiO.

\paragraph{Summary}
To summarize, epitaxial CoO(001) thin films have been grown with thicknesses of 2 and 4\,ML on Ag(001). For the 4\,ML thick film, antiferromagnetic (2$\times$1) LEED superstructures have been found at 231\,K, which disappear with increasing temperature at a Néel temperature of 285\,K. For both CoO(001) thin film systems, the image potential states are identified by 2PPE below the vacuum energy. Time-resolved UV pump and IR probe 2PPE finds ultrashort image potential lifetimes around 20 and 12\,fs for 2 and 4\,ML, respectively. For pump photon energies around the CoO band gap, we identify electron excitation into a flat Co $3d$-derived conduction band, for which a lifetime of 27\,fs at the conduction band minimum is determined for a 2\,ML CoO thin film on Ag(001).

\paragraph{Acknowledgments.} 
The authors thank R. Kulla for technical assistance. We gratefully acknowledge the financial support by the Deutsche Forschungsgemeinschaft (DFG, German Research Foundation) -- Project-ID 328545488 -- TRR~227, projects~A06.

\end{document}